# Micronucleus induction by 915 MHz Radiofrequency Radiation in *Vicia faba* root tips.

Bianca Gustavino[1]*, Giovanni Carboni[2], Roberto Petrillo[2], Marco Rizzoni[1], Emanuele Santovetti[2].

[1]Department of Biology and [2]Department of Physics, University of Rome Tor Vergata, Via della Ricerca Scientifica 1, 00133 Rome (Italy).

**ABSTRACT**

The mutagenic effect of radiofrequency electromagnetic field (RF-EMF) is evaluated by the micronucleus (MN) test in secondary roots of *Vicia faba* seedlings. Root exposures were carried out with 915 MHz continuous wave (CW) radiation for 72h, at power densities of 25, 38, 50 W/m$^2$. The specific absorption rate (SAR) corresponding to the experimental exposures was measured with a calorimetric method and fall in the range 0.3-1.8 W/kg. Results show a significant increase of MN frequency up to ten fold, correlated with the increasing power densities values.

# 1. INTRODUCTION

In the last decades growing attention has been paid to understanding the potential health effects linked to radiofrequency radiation (RFR) exposure, due to the increasing use of mobile phones and wireless networks, in particular among young people and children. The main concern comes from the possible carcinogenic effects related to the RF electromagnetic fields (RF-EMFs) emitted by these devices, which mainly operate in the range from 800 to 2500 MHz.

On the basis of a considerable number of investigations, especially of epidemiological data, indicating a direct association of human exposure to radiofrequency (RF) electromagnetic fields (EMF) with brain cancer development, the International Agency for Research on Cancer (IARC) have classified RF-EMF as possible human carcinogens, i.e., group 2B [1].

More recent studies have provided further data supporting the hypothesis of a causative effect of mobile phone RF emission on brain cancer development [2-5]. In addition, other malignant or benign tumors have been reported to be linked to the use of mobile phones [6; see also 7], such as non-Hodgkin's lymphoma [8], head and neck tumors [9] and testicular cancer [10]. In contrast with these, negative results from epidemiological data have been reported [11, 12] and criticism cannot be ignored about the interpretation of positive results [13, 14]. However, the uncertainty to consider these radiation as a potential risk factor for human health especially comes from meta-analysis studies [15-18].

Because of the causative link between cancer development and mutation induction, studies on the capability of RFR exposure to induce DNA damage and mutations are of primary relevance.

Conflicting results have been obtained, as in the case of a cell-type dependent effect detected after exposure of several mammalian cell types to levels of RFR exposure in the range of those of cellular phones [19]. Negative results from cytogenetic analyses and/or genotoxicity tests (namely Comet assay, γH2AX foci) are described in human fibroblasts and leukocytes [20, 21], human amniotic cells [22] and human peripheral lymphocytes [23]. Positive effects have been reported for DNA damage induction in exposed rats [24], mouse cell lines [25] and human trophoblasts [26], while increased frequencies of micronuclei, chromosomal aberrations and aneuploidy have been detected in several mammalian cell systems, including human cells, after exposure to RFR [27-29]. Effects of RFR exposure on other biological species have been also investigated, indicating the induction of DNA fragmentation and apoptosis in insects [30, 31] and of DNA fragmentation in conjunction with antioxidant stress response in earthworms [32].

In plant systems the alteration of cell cycle progression associated to oxidative stress in *Vigna radiata* [33] and the induction of cytogenetic effects in terms of chromosomal aberrations and micronuclei have been reported both in *Allium cepa* [34, 35] and in *Zea mays* [36].

Due to the limited information on mutagenic effects of RFR in plants, the present work analyses the mutagenic effects, in terms of micronucleus induction, in *Vicia faba* root tips after a 72h exposure to 915 MHz continuous wave radiation, at three different values of power densities (25, 38, 50 W/m$^2$), in our cases corresponding to SAR values included in the range of 0.3 and 1.8 W/kg.

The *Vicia faba* root tip micronucleus test is one of the most employed plant mutagenesis test because of its sensitivity to a wide variety of mutagenic compounds [37] and to extremely low doses of X-rays [38]. It has been used on various types of contaminated materials [39-41] and recently standardized by AFNOR, the French member organization of ISO. Proposals for protocol standardization have been also done [42].

## 2. MATERIALS AND METHODS

### 2.1 Plant material and germination

*Vicia faba* (broad bean) seeds were stored at 4 °C under dry conditions until use. Before experimental exposure seeds were soaked overnight in tap water, then placed in a thermostatic cabinet for germination, at 20 °C in a moistened atmosphere, in the dark. After 4 days tips of the primary roots were removed (approximately 5 mm from the distal end) to promote secondary root growth; then seedlings were settled in 115x52 mm plastic containers filled with tap water (about 200 ml), placing seeds on a plastic grid in order to hold them over the water surface. Seedlings were left in the cabinet under the above mentioned conditions for further 4 days. After this period secondary roots had appeared and were used for RFR exposure.

### 2.2 TEM cell, dose metrics and exposure conditions.

For RFR exposure a transverse electromagnetic (TEM) cell [43] was used (figure 1, a - b). A TEM cell is a waveguide in which the electromagnetic field propagates approximately as a plane wave, with the electric field perpendicular to the cell plates. For the present experiments an open TEM cell was built in our laboratory using copper clad glass-epoxy plates, according to Satav and Agarwal [44].

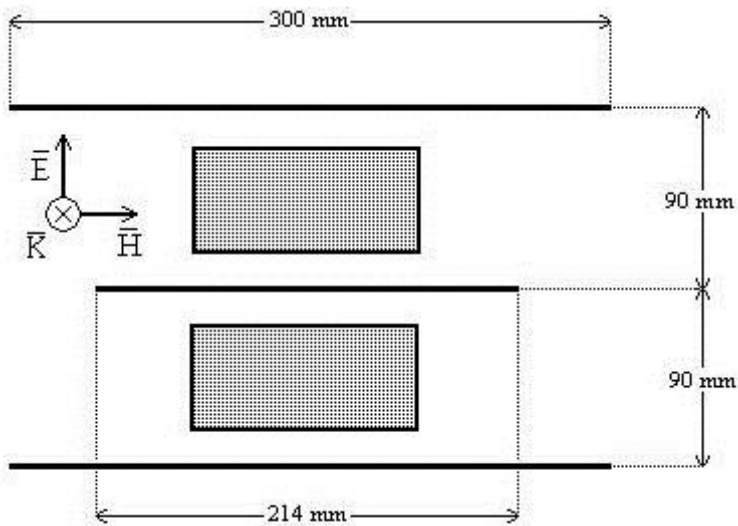 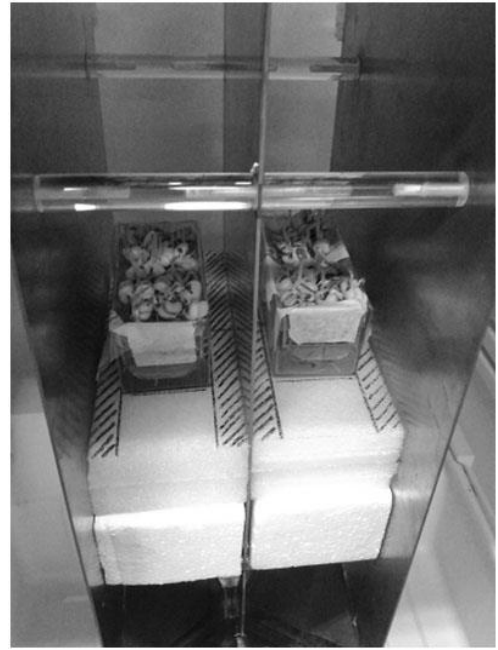

**Figure 1**. Transverse electromagnetic cell. [**a**] Schematic representation of the TEM cell (top view) in which the electric field (E), the magnetic field (H) and the wave (k) vectors are indicated. The gray rectangles represent the plastic box in which *Vicia faba* seedlings are placed for exposure. [**b**] An image of the TEM cell positioned inside the thermostatic cabinet is shown (front view).

The exposure setup is schematically shown in Fig. 2. The RF CW signal at 915 MHz, produced by a signal generator (Agilent E4420B), was sent to a power amplifier (Mini Circuits ZHL-5W-2G S+) and then to the TEM cell. The cell input power was monitored with an analog RF wattmeter (DAIWA CN-801), previously calibrated against a precision power meter (Gigatronics 8542C). The cell was terminated externally on its characteristic impedance to avoid any reflected signal in the system.

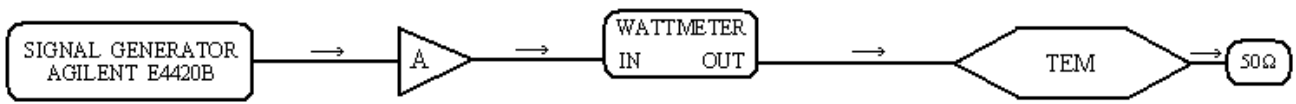

**Figure 2**. Exposure setup. The Agilent E4420B signal generator feeds an RF amplifier (Mini Circuit ZHL 5W 2G S+); the output is measured by an RF wattmeter (DAIWA CN-801), and sent to the TEM cell where the seedlings are exposed.

The *Vicia faba* seedlings, settled in two rectangular plastic containers filled with tap water (200 ml, corresponding to a level h= 35 mm), were positioned inside the TEM cell. The whole system was kept inside a thermostatic cabinet at a constant temperature of 20 $^0$C, in the dark (figure 1.b).

Each plastic container, with about 20 germinated seedlings of *Vicia faba*, was placed in the centre of one of the two compartments of the TEM cell that was vertically positioned inside the thermostatic cabinet. In this system the e.m. plane wave generated inside the TEM cell invests the exposed sample with the wave vector perpendicular to the water surface and the electric field vector parallel to the short edge of the rectangular container (figure 1.a: k- and E- vectors, respectively).

The experiment was carried out at three levels of RFR exposure at 915 MHz; the measured input powers entering the TEM cell were 1.5±0.1, 2.3±0.1 and 3.0±0.2 W, respectively. Sham exposure was carried out in the same TEM cell, without the transmission of RFR (power "off").

The electric field strength (V/m) of the plane wave propagating inside the cell is evaluated according to Satav and Agarwal [44] and Crawford [43]:

$$E = \frac{\sqrt{P_i \cdot Z_c}}{b/2} \qquad [1]$$

where E is the electric field, $P_i$ (W) is the power entering the cell, measured by the RF wattmeter, $Z_c$ is the characteristic impedance of the cell (in our case $Z_c$ = 50Ω) and b/2 is the distance between the inner and one of the outer plates of the TEM cell. The uncertainty is 0.5 dB.

The power density (*I*, intensity: W/m$^2$) is evaluated on the basis of:

$$I = \frac{E^2}{Z_0} \qquad [2]$$

where $Z_0$ is the characteristic impedance of the mean where the wave propagates (in air: $Z_0$ = 377 Ω).

The three exposure levels chosen for the experiment corresponds to the electric field strength of 98±5, 120±7, 138±8 V/m, and to the power density of 25±3, 38±4, 50±5 W/m$^2$ respectively.

The mean specific absorption rates (SAR) of the exposed water volumes were evaluated by a series of separate calorimetric measurements; the thermal transients following the RF exposures were analyzed, and the corresponding three SAR values for our exposure levels were found to fall in the ranges of 0.3–0.9, 0.5–1.4, 0.6–1.8 W/kg, respectively.

Samples of approximately 20 germinated seedlings of *Vicia faba* were used for each experimental point, including negative (sham) and positive controls. The highly mutagenic herbicide, maleic hydrazide (10$^{-4}$M water solution) was employed as positive control, exposing roots for 4h followed by a 68h recovery time. Seedlings were exposed for 72h to RF-EMF after which roots were immediately fixed.

**2.3 Slide preparation and cytogenetic analysis**

After exposure to RF-EMF (72h), secondary roots were excised, fixed in Carnoy solution (25% acetic acid:75% absolute ethanol, v:v) for 30 min; roots were then transferred in a fresh fixing solution and stored overnight at +4 °C. After Feulgen staining, root tips were squashed onto pre-cleaned slides in 45% acetic acid and permanently mounted in Eukitt (FoLabo, Italy). Micronucleus (MN) frequencies were calculated over 75000 cells per experimental point, where blind microscopic analysis was carried out by different operators, on the basis of 5000 cells scored per root tip, 5 tips/experimental point/operator. Only proliferating cell populations were considered for MN frequency analysis, on the basis of a contemporary mitotic index estimation of each root tip under study, and a minimum value of 2% mitotic cells was accepted for MN counting.

## 2.4 Statistical analysis

The statistical analysis was performed using *Graphpad Instat* software. Difference between the means were determined using the Mann-Whitney non parametric test. Values of P<0.05 were considered significantly different from sham exposure (negative control).

## 3. RESULTS AND DISCUSSION.

Micronucleus frequencies and the corresponding power density values are summarized in Table I.

**Table I.** Mean micronucleus (MN) frequencies (expressed per 5000 cells) detected in *Vicia faba* root tip cells after 72 hour exposure to 915 MHz RF Radiation. Values of power density (W/m$^2$) for the three levels of exposure are also shown.

| Treatment | Power Density (W/m$^2$)$^a$ ± SE | Mean MN Frequency $^b$ ± SE |
|---|---|---|
| sham | 0 ± 0.2 | 2.67 ± 0.52 |
| RF-EMF (915 MHz) | 25 ± 3 | 3.73 ± 0.62 |
|  | 38 ± 4 | 8.0 ± 1.0*** |
|  | 50 ± 5 | 20.2 ± 2.3*** |
| Positive control MH 10$^{-4}$M | 0 ± 0.2 | 37.5 ± 2.9*** |

$^{(a)}$ Values calculated with Equation [2].

$^{(b)}$ Mean MN frequencies are calculated on 15 tips/experimental point, 5000 cells/tip.

***=p<0.0001 (Mann-Whitney test).

The extremely significant difference of MN frequencies between maleic hydrazide and sham exposed *Vicia faba* cells indicates the sensitivity and reliability of this test system in the present experiment. A remarkable and extremely significant increase of MN frequencies is found for exposures to power densities of 38 and 50 W/m$^2$ (P<0.0001) compared to the sham value, which correspond to estimated SAR values falling in the ranges of 0.5-1.4 and 0.6-1.8 W/kg, respectively. It is noteworthy that the MN frequency detected at the highest power density is comparable to that induced by an X-ray irradiation at dose falling in the range between 8 and 12 cGy in the same biological system [38]. This suggests that RF EMF exposures may induce a strong mutagenic effect, in terms of both clastogenic and aneugenic effects [45] at relatively high values of power density.

A continuous and prolonged exposure was chosen in order to achieve a maximum yield of micronucleus frequency because an equilibrium frequency of micronuclei is reached at this exposure/fixation time, between newly induced micronuclei, arising at a constant rate from continuous exposure, and disappearance of old ones by dilution and/or disruption [46-48].

The observed increase of micronucleus frequency induced by RF-EMF exposure can hardly be attributable to thermal effect, because *Vicia faba* seedlings grown at 30°C did not show greater micronucleus frequency compared to those grown at 20°C, in the absence of mutagenic agents [48]. Similar conclusions have been made for the earthworm *Eisena fetida* exposed to 900 MHz EMF, excluding hyperthermia as a possible cause of the observed genotoxic effect [32].

Some comparisons can be made with results obtained in plant systems on mutagenicity end-points. As far as micronuclei are concerned our data can be compared with those obtained in *Allium cepa* [35], in which a remarkable increase in micronucleus frequency was observed with increasing exposure times to 890-915 MHz of RF-EMF, where frequencies of chromosomal aberration and aberrant mitoses also followed the same pattern.

Data from other experiments on *Allium* roots, in which RF exposure was carried out in a TEM cell, showed an increase of mitotic anomalies and chromosomal aberration frequencies at different electric field strength and frequency (400 and 900 MHz) values [34].

Exposures of *Zea mays* to 900 MHz, performed during different stages of germination, produced an increase of aberrant mitoses in exposed samples with respect to control ones [36]. Experimental exposure of *Lens culinaris* carried out at 1800 MHz, either before or during seed germination, led to an increased frequency of abnormal mitoses in exposed root cells [49].

Another set of research concerns the oxidative stress linked to genotoxicity. Studies on the induction of oxidative stress in plants by RF exposure have shown its influence in *Lemna minor*, both in terms of peroxidase activity [50] and of lipid peroxidation, hydrogen peroxide content and enzyme activity modulation [51] at 400 and 900 MHz. Results obtained from 900 MHz RF exposure of *Vigna murata* also found a significant upregulation of scavenging enzyme activities [33].

Our findings can be also viewed in a wider context and compared to those obtained in other biological systems, mainly mammalian cells, in which many authors found a lack of evidence about the induction of micronuclei by RF exposure, as reported in the meta-analysis studies [16-18]. Nevertheless, the induction of micronuclei by RF exposure has been shown in human fibroblasts [28], in exfoliated cells of exposed human individuals [52], in rat brain cells after *in vivo* exposure [53], in a brain cell culture model [29] and in circulating erythrocytes of rats [54]. Aneuploidy was also shown to be induced by RF exposure in human peripheral blood lymphocytes [27].

A genotoxic and preclastogenic effect induced by RF exposure has been shown in coelomocytes extracted from exposed *Eisenia fetida* through the alkaline Comet assay [32], in mammalian cell lines, namely rat brain cells [ 24; 53], spermatocyte-derived mouse cells through the FPG-Comet assay [25], human trophoblasts through the alkaline Comet assay [26] (Franzellitti *et al.*, 2010) and human fibroblasts [28] (Schwarz *et al.*, 2008) and by DNA repair foci (53BP1) in human cells [55].

In addition, the observed capability of RF exposure to induce an adaptive response to genotoxic agents in cultured human lymphocytes [56, 57] and mice [58] suggests that they are capable to induce genotoxic effects.

It has been also demonstrated that the oxidative stress induced by exposure to RF takes place in several mammalian cell systems (for a review see [59]), such as mouse spermatocyte-derived cell line [25], rat blood and brain tissue samples [60], human lens epithelial cells [61] human lymphocytes [62], human neuroblastoma and rat fibroblasts [63] and primary cultured neurons [64]. It has also been shown to be induced by RF exposure in plants [50, 51] and in the earthworm *Eisenia fetida* [32]. The induction of oxidative stress by RF exposure has been proposed to be responsible for their genotoxic effects [65; 32, 34, 55].

The mutagenic effect detected in the present experiments are in agreement with results obtained on all plant systems exposed to RF in which cytogenetic end-points are used. Moreover, the remarkable effect on MN induction by RF exposure, compared to the contradictory results obtained in mammalian cells, can be due to the much higher sensitivity of the MN test in *Vicia faba* root tips compared to MN test in mammalian cells [38].

# 4 Acknowledgments

Authors are sincerely grateful to Mr Giovanni Paoluzzi of the Department of Physics for providing precious technical support in setting up the exposure system. They also acknowledge the help of Dr. Walter Ciccognani of the Department of Electronic Engineering.